# Non-Hermitian Anomalous Scaling Engineering

Shulin Wang[1], Jiawei He[1], Zhiyuan Yang[1], Stefano Longhi[2,3,*], Peng Xue[1,†]

[1]*Key Laboratory of Quantum Materials and Devices of Ministry of Education, School of Physics, Southeast University, Nanjing 211189, China.*

[2]*Dipartimento di Fisica, Politecnico di Milano, Piazza Leonardo da Vinci 32, I-20133 Milano, Italy.*

[3]*IFISC (UIB-CSIC), Instituto de Fisica Interdisciplinar y Sistemas Complejos, E-07122 Palma de Mallorca, Spain.*

Corresponding authors:

[*]stefano.longhi@polimi.it

[†]gnep.eux@gmail.com

**Abstract**

Non-Hermitian systems exhibit anomalous scaling, a striking departure from conventional bulk laws, rooted in the non-Hermitian skin effect (NHSE). Here, we experimentally uncover this scaling and demonstrate its active control in a temporal photonic lattice. By tracking the real-time evolution of all eigenstates as system size varies, we directly observe scaling-driven spectral reshaping and eigenstate localization, revealing phenomena absent in Hermitian or NHSE-free lattices. In a Su-Schrieffer-Heeger lattice, scaling alone can trigger a non-Hermitian topological phase transition, with edge modes remaining protected. Crucially, Kerr interactions open the frontier of nonlinear non-Hermitian physics: weak nonlinearity accelerates or decelerates anomalous scaling, while strong nonlinearity suppresses it entirely. These results establish the first experimental platform for linear and nonlinear anomalous scaling engineering, paving the way for compact non-Hermitian devices and exploration of nonlinear and many-body non-Hermitian phenomena.

The non-Hermitian skin effect (NHSE) is a striking localization phenomenon in which all eigenstates accumulate at the system's boundaries, fundamentally breaking the conventional bulk-boundary correspondence that underlies much of traditional topological physics [1-8]. This unusual effect has attracted widespread interest, both for its deep implications in non-Hermitian and topological physics and for its potential in diverse applications, including topological light funneling, robust signal routing, enhanced classical and quantum sensing, and engineered photonic devices [9-16]. However, realistic devices are inevitably finite, making the understanding of scaling behavior crucial. Recent theoretical work predicts that the NHSE gives rise to anomalous scaling—a phenomenon exclusive to non-Hermitian systems, in which both spectra and eigenstate localization transform nontrivially with system size [17-26]. This scaling drives effects with no counterpart in any Hermitian or NHSE-free model, including the remarkable real-to-complex spectral transition and scale-free localization. Even more strikingly, anomalous scaling can act as a control knob that triggers non-Hermitian topological phase transitions [27,28]. Despite its significance as a unique signature of non-Hermitian physics, this behavior has remained experimentally out of reach, especially in optics, one of the most promising fields for applications. Beyond these linear predictions, an even broader challenge is emerging: understanding how anomalous scaling behaves in the presence of interactions and nonlinearities—an area central to the rapidly developing frontier of nonlinear and many-body non-Hermitian physics [29], yet with almost no experimental evidence to date. These gaps primarily stem from the difficulty of realizing highly reconfigurable platforms that enable simultaneous and precise control of both non-Hermitian and nonlinear effects.

Synthetic dimensions in photonics provide a versatile platform for probing non-Hermitian and nonlinear physics [30-35]. In particular, temporal photonic lattices constructed from two coupled fiber loops enable full control of inter-site couplings and complete access to the light field, including both spatial and temporal evolution [9,16,36-46]. Furthermore, nonlinearities—either intrinsic or artificially introduced—can be incorporated with high precision, making these lattices ideal for exploring nonlinear effects in non-Hermitian systems. Such capabilities position synthetic dimensions as a natural testbed for probing the unexplored interplay between non-Hermiticity, nonlinearity, and emergent collective dynamics, a direction increasingly recognized as foundational for next-generation non-Hermitian physics [29]. These platforms have already allowed the experimental observation of complex non-Hermitian effects, such as space-time topological events, non-Hermitian Dirac mass, and photon-photon thermodynamic processes [44-46].

In this work, we experimentally demonstrate, for the first time, anomalous scaling and its nonlinear control in the NHSE. Using a highly reconfigurable temporal lattice, we track the real-time evolution of all eigenstates across varying system sizes and directly observe scaling-induced transformations of spectra and eigenstate localization, covering both topologically trivial and nontrivial cases. Moving beyond the linear regime, we show that Kerr interactions offer a powerful route to manipulating anomalous scaling—accessing a domain in which interacting and nonlinear non-Hermitian dynamics remain largely uncharted experimentally. By introducing a weak Kerr nonlinearity, we accelerate or decelerate the anomalous scaling, whereas a strong nonlinearity leads to its complete suppression, highlighting the interplay between nonlinearity and NHSE anomalous scale localization. By enabling active control of anomalous non-Hermitian localization in finite-sized systems, these results unlock transformative pathways for compact nonlinear photonic devices and the exploration of rich non-Hermitian topological phenomena, marking an important step toward understanding nonlinear and ultimately many-body manifestations of anomalous scaling and establishing a foundation for practical non-Hermitian-based technologies.

We consider photonic quantum walks of optical pulses in a temporal mesh lattice, realized using two coupled fiber loops populated with circulating pulses [9,16,36-46], as illustrated in Fig. 1. The two loops possess a slight length difference and are connected through a variable optical coupler (VOC). When a pulse is injected, it is first split into two parts and then circulates in the different loops, enabling a time-multiplexing process. As the circulating number increases, the incident pulse is gradually extended into two long pulse trains, each residing in one of the two loops. Mapping the pulse evolution onto an $m$-$n$ mesh grid produces a synthetic lattice in the time domain, where $m$ is the circulating number, and $n$ is determined by the order of the pulse within the same circulation. Compared with the initial pulse, the pulse with a time delay of $mT + n\Delta T$ corresponds to the lattice site ($m$, $n$), where $T$ and $2\Delta T$ denote the average and difference of the two loops' time delays. At each lattice site, the coupling can be flexibly controlled by modulating the VOC via an arbitrarily programmed electric signal. Beyond the reciprocal couplings, applying loss and gain to the short and long loops enables the nonreciprocal couplings, since the pulse circulations in the short and long loops serve as the leftward and rightward couplings in the lattice. In addition, we insert optoelectronic feedforward circuits into the loops to generate Kerr nonlinearities of different strengths [16]. In such a temporal lattice, the pulse dynamics is described by the nonlinear map

$$\begin{cases} u_n^{m+1} = e^{-\gamma}\left[\cos(\beta_{n+1}^m)u_{n+1}^m + i\sin(\beta_{n+1}^m)v_{n+1}^m\right]e^{i\chi\left|e^{-\gamma}\left[\cos(\beta_{n+1}^m)u_{n+1}^m + i\sin(\beta_{n+1}^m)v_{n+1}^m\right]\right|^2}, \\ v_n^{m+1} = e^{+\gamma}\left[i\sin(\beta_{n-1}^m)u_{n-1}^m + \cos(\beta_{n-1}^m)v_{n-1}^m\right]e^{i\chi\left|e^{+\gamma}\left[i\sin(\beta_{n-1}^m)u_{n-1}^m + \cos(\beta_{n-1}^m)v_{n-1}^m\right]\right|^2}, \end{cases} \quad (1)$$

where $u_n^m$ and $v_n^m$ are the pulse amplitudes in the short and long loops normalized by the incidence, $\beta_n^m$ determines the coupling ratio of VOC at the site $(m, n)$ in the form of $\sin^2(\beta_n^m)/\cos^2(\beta_n^m)$, $\gamma$ is the gain and loss parameter, and $\chi$ is the effective Kerr coefficient of the feedforward circuit.

We first focus on the linear propagation regime ($\chi = 0$), with the coupling of the VOC set at zero, $\beta = 0$, in the bulk. To investigate the scaling effect, we consider a finite lattice. In the even and odd circulation steps, the sets of lattice sites are $\{-N, -(N-2), ..., N-2, N\}$ and $\{-(N-1), -(N-3), ..., N-1, N+1\}$, where $N$ is an even integer. In addition, we introduce a coupling of $\beta_b$ between the sites $-(N-1)$ and $N+1$ in odd steps to realize the boundary coupling, i.e., the generalized boundary condition (GBC). Due to the discrete translational invariance of the lattice in the $m$ direction, the pulse amplitude evolution in the lattice can be expanded into $(u_n^m \; v_n^m)^T = (U \; V)^T z^{n/2} e^{i\theta m/2}$, where $(U \; V)^T$ is the eigenvector, $z$ is the generalized Bloch factor [1-8], equals to $\exp(iQ)\exp(h)$ with $\exp(iQ)$ and $\exp(h)$ being the phase factor and amplitude gain factor, and $\theta$ is the quasienergy or propagation constant. By inserting the Ansatz into the pulse evolution equation (1), we obtain the quasienergy band structure

$$\cos\theta = \left(ze^{-2\gamma} + z^{-1}e^{2\gamma}\right)/2. \quad (2)$$

For a given $\theta$, there exist two generalized Bloch factors, i.e., $z_1$ and $z_2$, satisfying $z_1 z_2 = \exp(4\gamma)$. The eigenstate with $\theta$ is thus a superposition of two non-Bloch states. Considering the GBC, the available quasienergies are given by

$$\theta = -Q - 2i\gamma + i\ln|Z|/(N+1), \quad (3)$$

in which $Z = z_2^{N+1} = \cos(\beta_b)\left[1+e^{4(N+1)\gamma}\right]/2 \pm \sqrt{\cos^2(\beta_b)\left[1+e^{4(N+1)\gamma}\right]^2 - 4e^{4(N+1)\gamma}}/2$, $Q = (\text{Arg}Z+l\pi)/(N+1)$, and $l = -N, -(N-2), ..., N-2, N$. By further calculating the superposition coefficients of the two non-Bloch modes, we can obtain the corresponding eigenstates. From the spectrum, the energy gain factors of the eigenstates can be calculated as

$$G = \exp\left[-2\operatorname{Im}(\theta)\right] = \exp\left[4\gamma - 2\ln|Z|/(N+1)\right]. \qquad (4)$$

For a short lattice chain with $N \to 0$, the quasienergies are close to $\theta = -Q$, where $Q = (\pm\pi/2 + l\pi)/(N+1)$. Such a solution is identical to the case with open boundary condition (OBC). For a long lattice chain with $N \to \infty$, the quasienergies approach $\theta = -Q \pm 2i\gamma$, where $Q = l\pi/(N+1)$. This result is the same as the case with periodic boundary condition (PBC). See Supplementary Material Sec. I for more details on the above derivation.

Figure 2(a) illustrates the evolution of the GBC spectrum with increasing lattice sizes, together with comparisons to the OBC and PBC spectra of an infinite chain. Here, the non-Hermitian parameter and boundary coupling are chosen as $\gamma = 0.2$ and $\beta_b = \arccos(0.02)$. One sees that the spectra of short chains are completely real and overlap with that of the OBC case. However, as the system size increases, the imaginary part emerges and becomes increasingly pronounced. For a sufficiently large lattice size, the GBC spectrum nearly coincides with the PBC spectrum. The energy gain or loss of the eigenstates is the most prominent manifestation of this spectrum evolution. As shown in Fig. 2(b), for $N = 8$ the eigenstates remain conserved during propagation. With $N$ increased to 12 and 16, gain and loss appear, and their strengths grow with the system size. In addition to the spectrum, the eigenstate profiles also display a scale-dependent property. Figure 2(c) depicts the inverse participation ratio (IPR) of the averaged eigenstate profile varying with $N$. As the system size increases, the IPR of the GBC case first overlaps with that of the OBC situation and then gradually approaches that of the PBC case. Moreover, the IPR gradually transitions from a high value to nearly zero, indicating that the eigenstates change from localized skin states to extended states. Nevertheless, in conventional cases such as a Hermitian lattice, the spectrum merely becomes denser as the lattice size increases, and the eigenstates remain extended throughout. The above scaling behavior of the NHSE thus stands in stark contrast to the traditional scenarios.

To demonstrate the scale-dependent properties of NHSE in the linear regime, we experimentally establish the dual-loop circuit based on optical communication systems. The two fiber loops have a length of ~8 km and a length difference of ~0.4 km, as detailed in Supplementary Material Sec. II. Additionally, the closed lattice chain with the GBC is effectively emulated using an open chain, as described in Supplementary Material Sec. III. In our experiments, we precisely excite all the eigenstates for $N = 8$, 12, and 16 and observe their evolution over 10 steps. By recording the gain or loss of the eigenstates during evolution, the energy gain factors are successfully measured. As shown in Fig. 2(b), the measured results agree well with the theoretical prediction. To gain deeper insight into the scaling behavior, we display the evolution of several typical eigenstates in Fig. 2(e), with the simulation in Fig. 2(d) serving as a comparison. For $N = 8$, the eigenstate with an index of $J = 1$ exhibits almost conservative propagation, and the measured pulse intensity evolution coincides well with the simulation. For $N = 12$ and $J = 1$, the eigenstate continuously acquires a slight energy gain during propagation. For $N = 16$ and $J = 1$, the gain effect is significantly enhanced due to the enlargement of the spectrum. Conversely, for $N = 16$ and $J = 2$, a dramatic loss occurs during mode evolution, as the corresponding quasienergy has a considerable positive imaginary part. One also sees that the localization of the eigenstate profiles degrades as $N$ increases from $N = 8$ to 16, which can be regarded as the scaling behavior of eigenstates. It should be noted that the excited eigenstates in our experiments maintain nearly constant intensity profiles during evolution, demonstrating the effectiveness of our eigenmode preparation setup.

Intriguingly, scaling the lattice can even trigger a non-Hermitian topological phase transition. To study topological effects, we apply Floquet modulation to the coupling of the lattice and construct an effective Su-Schrieffer-Heeger model. At the even and odd steps, the coupling in the bulk is fixed at $\beta_1$ and $\beta_2$, respectively. We first concentrate on the OBC and PBC cases, which can be regarded as two specific examples of GBC. The OBC and PBC phase diagrams, i.e., the Zak phases as functions of $\beta_1$ and $\beta_2$, are derived and shown in Figs. 3(a) and 3(b), with the detailed derivation provided in Supplementary Material Sec. IV. One can see that the two phase diagrams are quite similar, but the OBC one exhibits a wider topologically non-trivial region. Consider a carefully chosen set of couplings (e.g., $\beta_1 = \pi/10$ and $\beta_2 = 0$), which lie in the topologically non-trivial and trivial regions in the OBC and PBC phase diagrams, respectively. Figure 3(c) shows the calculated GBC spectrum. For small system sizes, topological edge states with real quasienergies exist near $\pm\pi$ and are separated from the bulk band. As the size increases, the edge states gradually approach the bulk band and eventually vanish. Similar to typical topological cases, the average IPR of the edge states is higher than that of bulk states, as depicted in the inset of Fig. 3(c). These phenomena indicate a scaling-induced non-Hermitian topological phase transition. Traditionally, the topological phase is determined by the band structure of an infinite chain. Following this paradigm, the temporal lattice with the current couplings should always be topologically trivial, since the GBC spectrum of an infinite temporal lattice is identical to the PBC one. However, at small system sizes, the GBC spectra approach the topologically non-trivial OBC one, thereby enabling the existence of topological edge states. In the experiments, we prepare the edge state near $-\pi$, i.e., the mode with $J = 1$, and monitor its evolution. As illustrated in Fig. 3(e), the state evolves stably and displays a more localized intensity profile compared to the bulk states. The extraordinary topological edge state at a small size is hence demonstrated. For a set of couplings that locate at the topologically non-trivial region in both OBC and PBC phase diagrams (e.g., $\beta_1 = \pi/3$ and $\beta_2 = 0$), topological edge states exist across different lattice sizes [Fig. 3(d)]. Due to robust topological protection, the edge states can be almost immune to the scaling effect when they are far away from the bulk band. The measured edge state evolution for $N = 10$ and 16 evidently verifies such independence from the scale, where the two edge states exhibit nearly identical intensity patterns [Figs. 3(f) and 3(g)].

Recent pioneering works have revealed that nonlinearity can play a key role in manipulating the NHSE [16,47-50]. However, how nonlinearity can be used to engineer and control the *scaling behavior* associated with the NHSE—particularly the emergence of anomalous or non-standard scaling—has remained unexplored, both theoretically and experimentally. Here, we introduce and experimentally demonstrate a nonlinear control mechanism that directly tunes the anomalous scaling of linear skin modes, revealing a qualitatively new dimension of NHSE physics. Remarkably, even under very weak nonlinearities, the nonlinear eigenstates depart from the expected linear behavior in a quite sensitive manner, enabling precise manipulation of the scaling properties.

Under a very weak nonlinearity, the nonlinear quasienergies are expected to closely resemble those of linear systems. By using the self-consistent iterative algorithm [51] and feeding it with the linear eigenstates, we can numerically solve the nonlinear eigen equation, as discussed in Supplementary Material Sec. V. As shown in Fig. 4(a), where the coupling angle is $\beta = \pi/3$ in the bulk, the quasienergies at $\chi = \pm 0.05\pi$ nearly overlap with the linear ones because the weak Kerr nonlinearity introduces only minimal additional phases during propagation. However, due to the effects of nonlinear Kerr self-focusing and self-defocusing [38,52,53] at $\chi = \pm 0.05\pi$, the corresponding scaling behavior of the eigenstates can be efficiently decelerated and accelerated, as reflected by the IPR of the states [Fig. 4(b)]. It should be noted that the nonlinear eigenstate here originates from the $(N+1)$-th linear mode, which possesses a real quasienergy and can acquire intensity-dependent nonlinear phases in a constant manner. Figures 4(c) and 4(d) illustrate the measured evolution of nonlinear eigenmodes at $N = 10$ and 16 for $\chi = \pm 0.05\pi$. Compared to the linear case, the localization can be enhanced for $\chi = 0.05\pi$ and weakened for $\chi = -0.05\pi$, while the scaling behavior persists. For a high nonlinearity, the interplay of strong nonlinear self-trapping and NHSE gives rise to skin solitons with pronounced localization [16]. The nonlinear eigenstates can localize at almost a single site and thus become independent of the lattice scaling [Figs. 4(a) and 4(e)]. As shown in Fig. 4(f), the nonlinear eigenstates remain almost constant from $N = 10$ to 16 and exhibit significantly higher localization compared to their linear counterparts. Hence, the strong Kerr nonlinearity leads to a breakdown of the anomalous scaling behavior of the NHSE owing to nonlinear (soliton-like) localization.

In conclusion, we have realized the first direct experimental demonstration of non-Hermitian anomalous scaling with nonlinear control. Using a reconfigurable temporal photonic lattice with boundary coupling, we show that increasing system size drives a striking spectral transformation from real to complex and induces pronounced, size-dependent reshaping of skin modes. This anomalous scaling enables a non-Hermitian topological phase transition in the Su-Schrieffer-Heeger configuration, with edge modes protected whenever spectrally isolated from the bulk. A central advance of this work is the experimental validation of nonlinear control over anomalous non-Hermitian scaling, previously unexplored. We show that even weak Kerr nonlinearities can accelerate or decelerate the scaling, providing a direct means to tune non-Hermitian transport and localization. At higher nonlinearities, we observe skin solitons, signaling a nonlinear breakdown of the scaling regime.

Our results establish anomalous NHSE scaling as an accessible and tunable physical resource, charting a route toward finite-sized non-Hermitian photonic devices whose topological, directional, and nonlinear functionalities can be programmatically engineered via system size, boundary design, and material nonlinearity. More broadly, this work lays the foundation for a new generation of scalable, reconfigurable, nonlinear non-Hermitian technologies with impact across photonics, metamaterials, and quantum platforms.

This work has been supported by the National Key R&D Program of China (Grant No. 2023YFA1406701) and the National Natural Science Foundation of China (Grant Nos. 92265209 and 62305122).

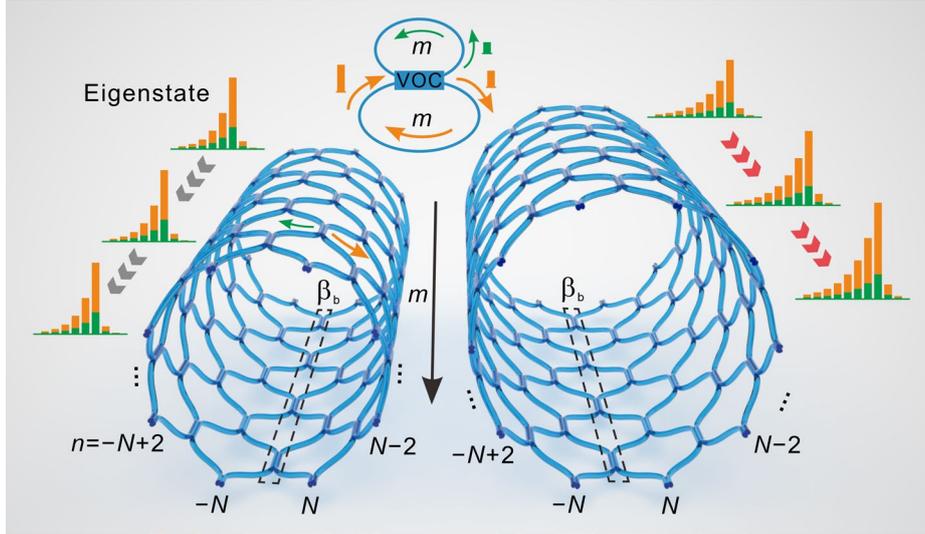

FIG. 1. Schematic of the temporal photonic mesh lattice originating from two coupled fiber loops (top inset). The two ends of the mesh lattice, i.e., the sites $n = -N$ and $N$, are connected through a boundary coupling $\beta_b$ to realize the GBC. The left (right) panel displays a short (long) lattice chain that supports conservative (amplified) eigenstate evolution.

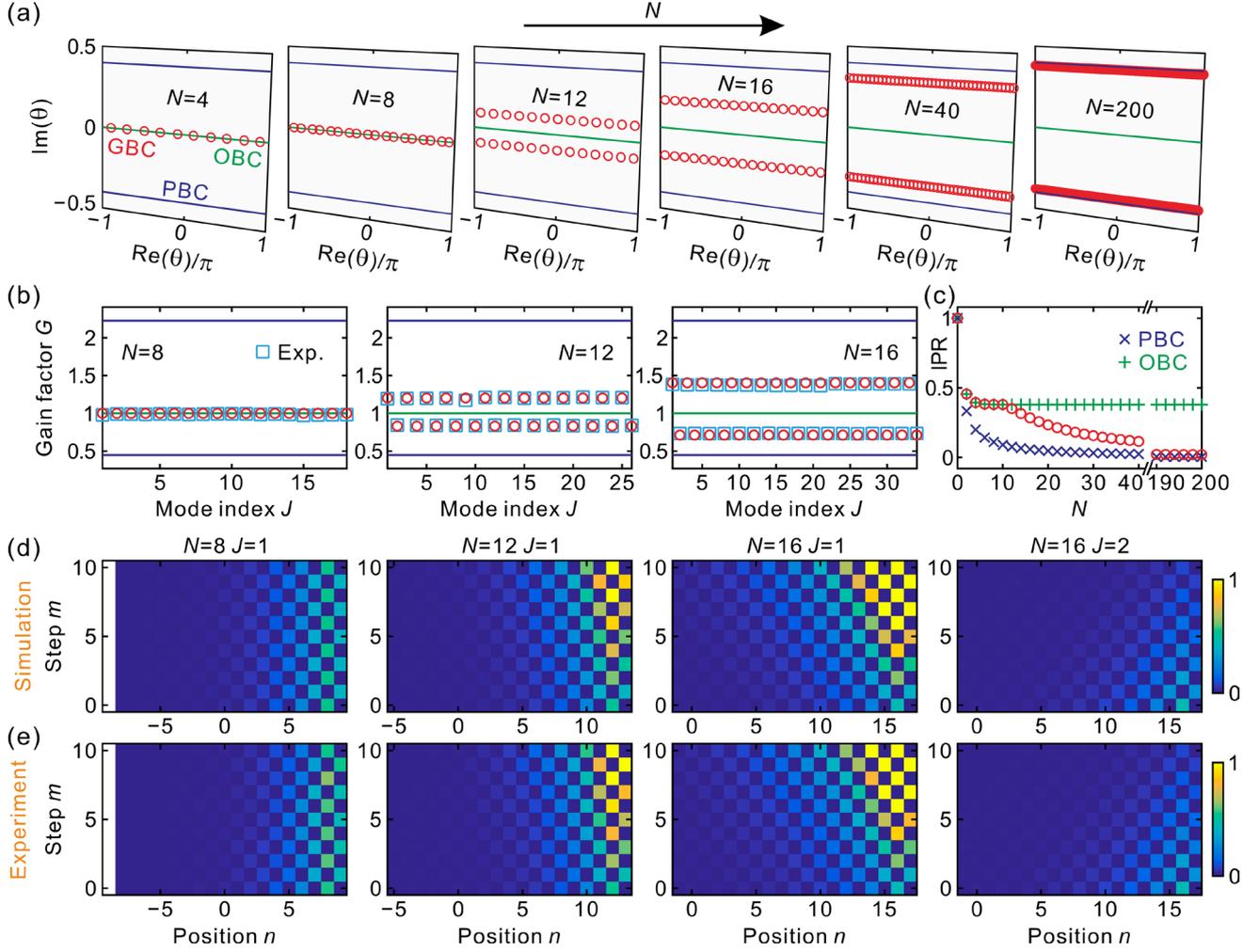

FIG. 2. (a) GBC spectra for $N$ = 4, 8, 12, 16, 40, and 200. The red circles represent the GBC spectra for finite lattice sizes, while the green and blue lines denote the OBC and PBC spectra of an infinite lattice chain, respectively. (b) Gain factors of light energy $G$ with respect to mode index $J$ for $N$ = 8, 12, and 16. The circles and squares correspond to the simulated and measured results, respectively. (c) Average IPR of eigenstates varying with $N$ under GBC, OBC, and PBC. (d) Simulated pulse intensity evolution for $N$ = 8, 12, and 16. (e) Corresponding experimental results.

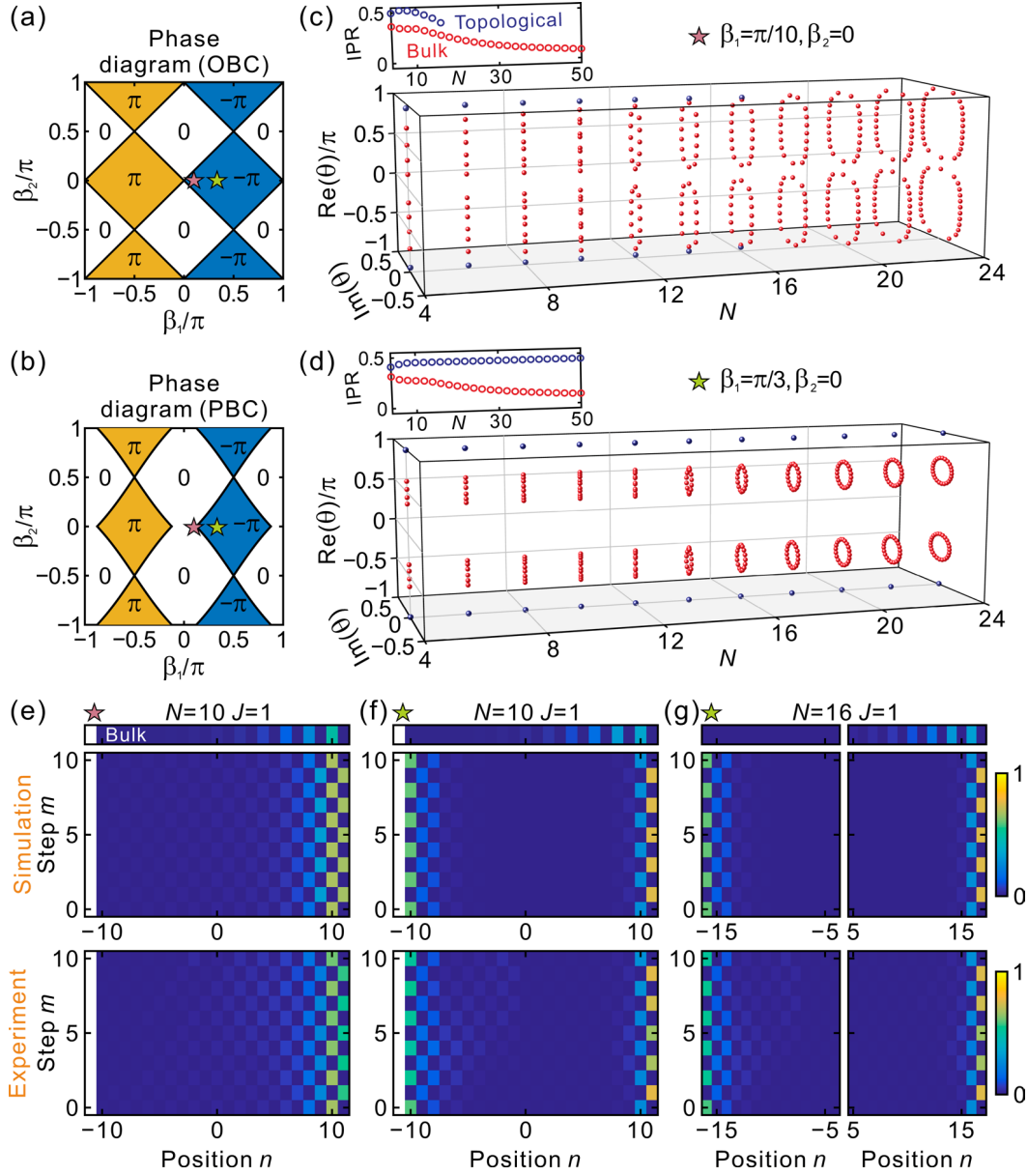

FIG. 3. (a), (b) Zak phase diagrams of the Su-Schrieffer-Heeger temporal mesh lattice under OBC and PBC, respectively. (c) GBC spectrum versus $N$ for $\beta_1 = \pi/10$ and $\beta_2 = 0$. The red and blue balls correspond to bulk and topological states, respectively. The inset depicts IPR varying with $N$. (d) GBC spectrum as a function of $N$ for $\beta_1 = \pi/3$ and $\beta_2 = 0$. (e) Simulated and measured intensity evolution of topological edge state ($J = 1$) at $N = 10$, $\beta_1 = \pi/10$, and $\beta_2 = 0$. The top bar is the average intensity profile of all bulk states. (f), (g) Simulated and measured intensity evolution of topological edge state ($J = 1$) at $N = 10$ and 16. The couplings are chosen as $\beta_1 = \pi/3$ and $\beta_2 = 0$.

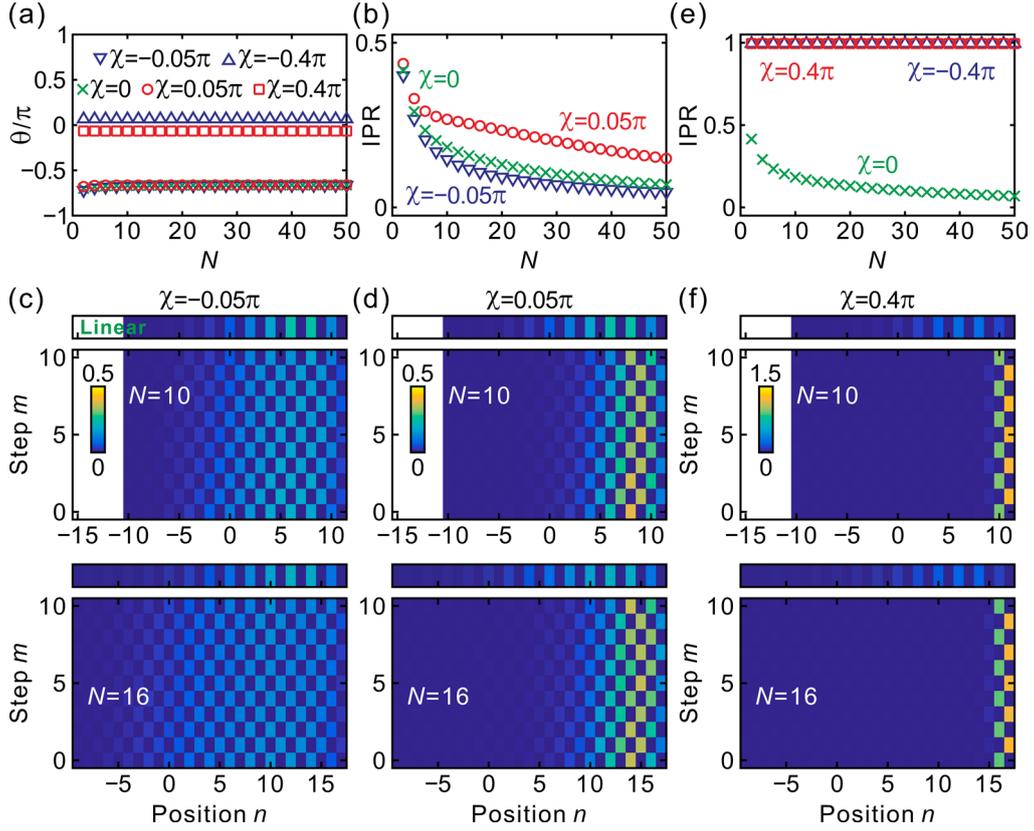

FIG. 4. (a) Quasienergy of nonlinear eigenstate versus $N$ for $\chi = \pm 0.05\pi$ and $\pm 0.4\pi$, together with the quasienergy of the $(N+1)$-th linear eigenstate ($\chi = 0$). (b) IPR versus $N$ at weak nonlinearity. (c), (d) Experimental intensity evolution at $\chi = -0.05\pi$ and $0.05\pi$. The upper and lower subplots correspond to $N = 10$ and 16, and the top bar within each subplot is the intensity profile of the eigenstate in the linear case. (e) IPR versus $N$ at strong nonlinearity. (f) Measured intensity evolution at $\chi = 0.4\pi$.